\documentclass[sigconf]{acmart}
\renewcommand\footnotetextcopyrightpermission[1]{} 
\usepackage{booktabs} 
\usepackage{todonotes}
\usepackage{balance}

\title{Deep Learning-based Concept Detection in vitrivr at the Video Browser Showdown 2019 -- Final Notes}

\author{Luca Rossetto$^1$, Mahnaz Amiri Parian$^{1,2}$, Ralph Gasser$^1$, Ivan Giangreco$^1$, \\ Silvan Heller$^1$, Heiko Schuldt$^1$}
\affiliation{%
  \institution{$^1$Department of Mathematics and Computer Science, University of Basel, Switzerland}
  \institution{$^2$Numediart Institute, University of Mons, Belgium}
}
\email{{firstname.lastname}@unibas.ch}

\begin{document}

\begin{abstract}
This paper presents an after-the-fact summary of the participation of the vitrivr system~\cite{rossetto2016vitrivr} to the 2019 Video Browser Showdown~\cite{cobarzan2017interactive}. Analogously to last year's report~\cite{rossetto2018competitivereport}, the focus of this paper lies on additions made since the original publication~\cite{rossetto2019deep} and the system's performance during the competition.
\end{abstract}

\maketitle


\section{Introduction}
vitrivr is an open-source multimedia retrieval stack which has -- together which its predecessor, the IMOTION system -- participated in the VBS five times so far~\cite{rossetto2015imotion,rossetto2016imotion,rossetto2017enhanced,rossetto2018competitive,rossetto2019deep} and was ranked $1^{st}$  for the second  time this year after it had already won the 2017 installment of the competition.

\begin{figure*}
\centering
\includegraphics[width=0.9\textwidth]{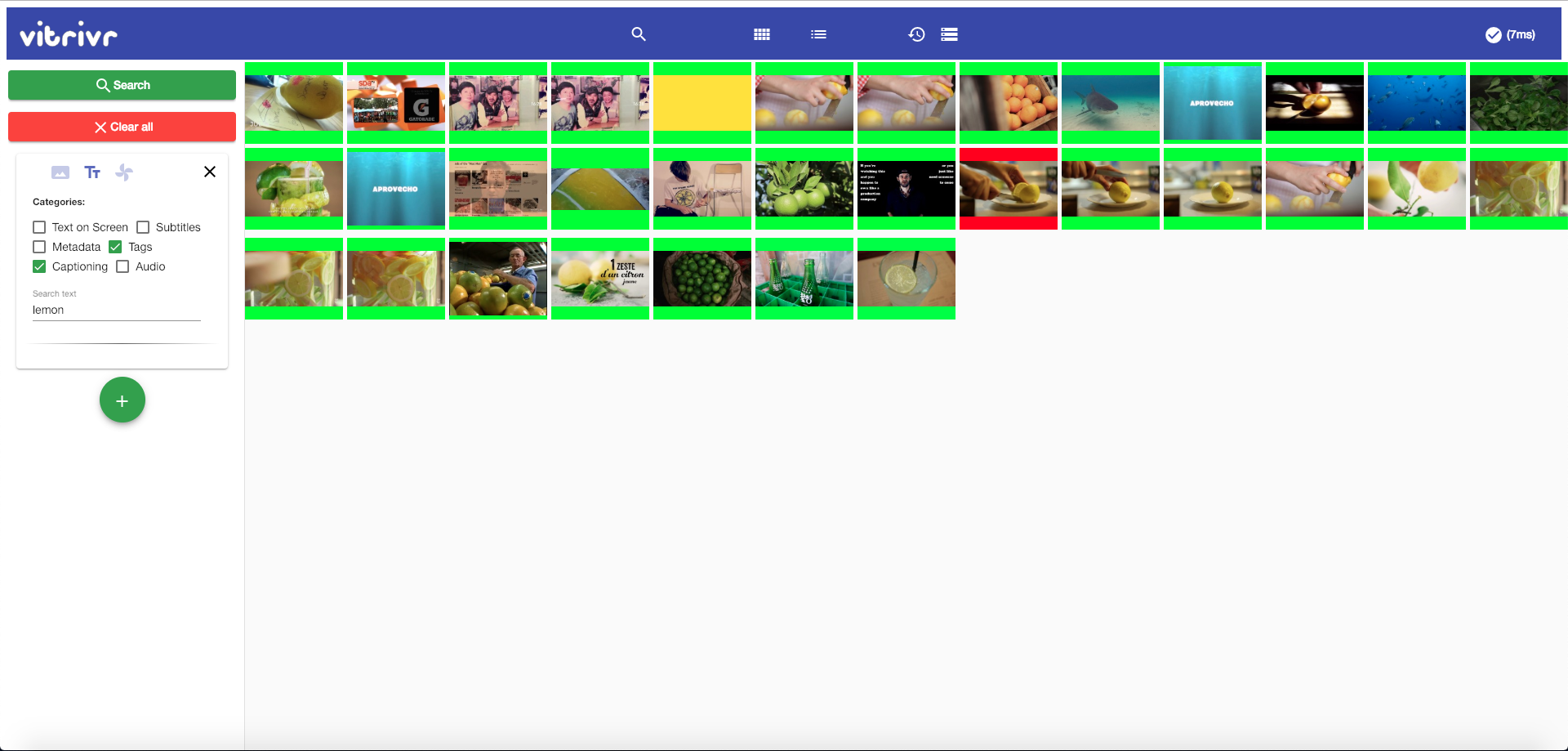}
\caption{Screenshot of the vitrivr user interface. It lists the results for an actual task that was part of the competition. The red item was successfully submitted to the server. }
\label{fig:screenshot}
\end{figure*}

This paper provides an addendum to the original publication~\cite{rossetto2019deep} by describing the additions made to the system that lacked documentation (Section~\ref{sec:additions}). Furthermore, we discuss the system's overall performance during the competition (Section~\ref{sec:performance}) as well as the lessons we have learned (Section~\ref{sec:lessons}). 

\section{Additions}
\label{sec:additions}
This section provides an overview of the additions that were made after the submission of the original publication~\cite{rossetto2019deep}. These additions were mostly concerned with efficiency and ease-of-use in a competitive setting such as VBS.


\subsection{Feature Extraction}

As mentioned in the original paper, we made use of various methods to extract or generate textual descriptions from the video content. In order to label actions that were performed during a video shot, we used the \emph{Inflated 3D Convolution I3D}\footnote{\url{https://github.com/deepmind/kinetics-i3d}} network pretrained on the \emph{Kinetics 600}\footnote{\url{https://deepmind.com/research/open-source/open-source-datasets/kinetics/}} dataset~\cite{kay2017kinetics}, which covers 600 different actions.

Since the new dataset~\cite{rossetto2019v3c}, in contrast to the previous one, does not include any ASR data, we generated the speech transcripts for the videos containing English dialogue using the Google Cloud Speech-to-Text API\footnote{\url{https://cloud.google.com/speech-to-text/}}.

\subsection{Collaborative Retrieval}

Since the setup during the competition called for two independent instances of the retrieval system, operated by one user each, we implemented a small coordination utility\footnote{\url{https://github.com/vitrivr/Collabordinator}} in order to facilitate collaborative retrieval. The user interface already had the option to highlight specific results with a user-definable color. In the new version, a result that has been submitted to the competition server would now also automatically be assigned a color. Figure~\ref{fig:screenshot} depicts an example of a submitted result in red. The coordination utility shares these color labels across several instances of the user interface, so one user can see what the other has already labeled or submitted. This functionality turned out to be especially useful during the Ad-hoc Video Search (AVS) tasks, since it greatly reduced redundancies in submitted results and did therefore allow for a more efficient use of the limited time available.

\subsection{Improvements of Interaction Efficiency}

In order to make the interaction with the system more efficient in a competitive setting, we made several minor improvements to the user interface which are briefly described as follows:

\begin{itemize}
    \item The Query by Sketch input was amended by a palette of frequently-used colors such as various skin tones.
    \item An additional filter step was added to remove retrieved results from the user interface. Filtering is possible by score, so as to exclude results below a certain score threshold, and by dominant color of a shot. The latter does for example enable the user to only show results that are predominantly `orange' or `yellow' in hue.
    \item Since we greatly increased the number of returned results, the display logic of the user interface was changed such that it would load the preview images on demand rather than all at once in order not to overly stress the browser running the UI. Additional optimizations were also made to better cope with the greatly enlarged result set.
    \item The user interface does now store the results from all previous queries. A dedicated menu lists the entire query history and allows a user to load any previous result without generating load on the retrieval back-end. The idea behind this was that in a competitive setting, the user should be able to go back to a successful query after having issued an unsuccessful one.
    \item We added the option to submit an item by holding down the `Shift'-key and clicking on it. Normally, clicking the result would open the retrieved video at the relevant point in time. This simple change improved the submission efficiency dramatically, especially during Ad-hoc Video Search (AVS) tasks. In previous versions, this functionality was hidden in a dedicated menu that would only appear when hovering a result item with the mouse.
\end{itemize}

\section{System use and Performance}
\label{sec:performance}

Even though vitrivr offers several different query modalities, both expert and novice users used text-based queries almost exclusively. This is not surprising, since in a time-constrained setting, textual input is quicker to produce than the visual alternatives. Especially for queries that contained either sufficient dialog or clearly visual text on screen, textual queries in the corresponding categories proved not only efficient but also highly effective. In the absence of such cues, the large number of labelled concepts and actions as well as the scene captions provided ample basis for textual search.

In this iteration of the competition, vitrivr performed well in all task categories. Out of the five, it was ranked first in three of them and was only outranked by the VIRET~\cite{lokovc2019viret} system in the other two. The overall placements for the different task types were as follows:
\begin{itemize}
\item 2$^{nd}$ place in textual KIS tasks (80/100 points)
\item 1$^{st}$ place in visual expert KIS tasks (100/100 points)
\item 1$^{st}$ place in novice visual KIS tasks (100/100 points)
\item 2$^{nd}$ place in expert AVS tasks (76/100 points)
\item 1$^{st}$ place in novice AVS tasks (100/100 points)
\item 1$^{st}$ place overall (91/100)
\end{itemize}

\section{Lessons Learned}
\label{sec:lessons}

In the following, we summarize some of the observations made during the competition:

\begin{itemize}
    \item Text is still a very effective and intuitive way to specify queries, especially for novice users.
    \item Visual Query-by-Sketch is not sufficiently selective for a collection of the size that is currently in use and does therefore (at least in a competitive setting) not justify the time required for query formulation and execution.
    \item The semantic sketch queries look promising but their execution time needs to be reduced in order to be useful in a competitive setting.
    \item In a time-constrained setting, browsing through a larger result set is more effective than trying to make the query more precise in order to reduce that set. 
    \item Despite its effectiveness, browsing can still lead to false-negatives as did happen in the 9$^{th}$ visual Known-Item Search (KIS) task, where the correct result was retrieved by the system but overlooked by both users.
    \item For Ad-hoc Video Search (AVS) tasks, efficient submission is just as important as effective retrieval. The added shortcuts for submission as well as the result synchronization were very helpful in that respect.
\end{itemize}

\section{Conclusion}
\label{sec:conclusion}

While vitrivr performed well in this edition of the Video Browser Showdown, there remain several improvements to be made. Large-scale video search is far from being a solved problem and variability in the outcomes of VBS over recent years show that in such a competitive setting, the final results are not an absolute measure for one system's performance with respect to another. Creativity in query formulation and to a certain degree plain luck also influence the final ranking of the systems. 

\balance 

\bibliographystyle{ACM-Reference-Format}
\bibliography{bibliography}
\end{document}